\documentclass[aps,twocolumn,superscriptaddress,floatfix,nofootinbib]{revtex4-2}

\usepackage[utf8]{inputenc}
\usepackage{amssymb}
\usepackage[dvipsnames]{xcolor}
\usepackage{amsmath}
\usepackage{graphicx}
\usepackage{appendix}
\usepackage{algorithmic}
\usepackage{float}
\makeatletter
\let\newfloat\newfloat@ltx
\makeatother
\usepackage{algorithm}
\usepackage{placeins}
\usepackage{bm}
\usepackage{lipsum}
\usepackage{listings}
\usepackage{tikz}
\usetikzlibrary{quantikz}

\begin{document}

\global\long\def\ket#1{\left|#1\right\rangle }%

\global\long\def\bra#1{\left\langle #1\right|}%

\global\long\def\linner#1#2{\left\langle \left.#1\,\right|#2\hspace{1.2pt}\right\rangle }%

\global\long\def\rinner#1#2{\left\langle \hspace{1.2pt}#1\left|\,#2\right.\right\rangle }%

\newcommand{\code}[1]{\texttt{#1}}

\title{Quantum Neural Networks for a Supply Chain Logistics Application}

\author{Randall Correll}
\affiliation{QC Ware Corp., Palo Alto, CA USA}
\author{Sean J. Weinberg}
\affiliation{QC Ware Corp., Palo Alto, CA USA}
\author{Fabio Sanches}
\affiliation{QC Ware Corp., Palo Alto, CA USA}
\author{Takanori Ide}
\affiliation{
AISIN CORPORATION, Tokyo Research Center, Chiyoda-ku, Tokyo, Japan
}
\author{Takafumi Suzuki}
\affiliation{Aisin Technical Center of America, San Jose, CA USA}

\date{\today}

\begin{abstract}

Problem instances of a size suitable for practical applications
are not likely to be addressed during the noisy intermediate-scale quantum (NISQ) period with (almost) pure quantum algorithms. 
Hybrid classical-quantum algorithms have potential, however,
to achieve good performance on much larger problem instances.
We investigate one such hybrid algorithm on a problem
of substantial importance: vehicle routing for supply chain
logistics with multiple trucks and complex demand structure.
We use reinforcement learning with neural networks with 
embedded quantum circuits.
In such neural networks, projecting high-dimensional feature vectors
down to smaller vectors is necessary to accommodate restrictions
on the number of qubits of NISQ hardware. However, we use
a multi-head attention mechanism where, even in classical machine
learning, such projections are natural and desirable.
We consider data from the truck routing logistics of a company in the automotive sector, and apply our methodology by decomposing
into small teams of trucks, and we find results
comparable to human truck assignment.

\end{abstract}

\maketitle

\section{Introduction}

In this work, we take explore a hybrid quantum-classical orthogonal neural network implemented in a reinforcement learning (RL) scheme to develop agents that can obtain good solutions in complex supply chain problems.  In the setting of commercial operations, computational
problems of substantial theoretical and practical difficulty regularly
arise.  Even a small improvement on the quality of solutions
to such problems can translate to a very substantial benefit.
One such problem, heavily studied in operations research,
is the vehicle routing problem \cite{dantzig1959truck, toth2002vehicle}.

In a previous effort \cite{poc3cl}, we build on the work of \cite{kool2018attention} by adding new techniques allowing for multiple trucks and for far more general requirements for trucks. While our model
is specially designed with Aisin Corporation's vehicle routing
problems in mind, our techniques for applying RL to a very
general class of routing problems apply widely.

Vehicle routing problems are NP-hard combinatorial optimization
problems for which there are numerous heuristic algorithms
which yield approximate solutions. Relatively recently,
there has been interest in solving such routing problems
using reinforcement learning (RL) \cite{bello2016neural}. 
In this context, a truck driving between nodes can be thought of as an agent
performing actions (selecting its next node to drive to)
in the environment of the supply chain. 

The advent of quantum computing promises potential speedups in computation for a variety of problems.  How well they apply to heuristic algorithms such as machine learning is an active area of research.  Here we adapt quantum orthogonal neural networks as part of an attention head mechanism in a reinforcement learning scheme.  The quantum circuit at the heart of the orthogonal neural network implemented in a supply chain logistics workflow is tested both using classical emulation and implementation on real quantum hardware.  The experiments yields results which are preliminary but promising for the implications for quantum-boosted algorithms and applications as the underlying quantum algorithms and quantum computing hardware mature.

\section{Quantum Orthogonal Neural Networks}
\label{sec:quantum-onn}

Orthogonal neural networks (ONNs) are a type of
neural network where linear layers are constrained to be orthogonal matrices \cite{li2019orthogonal}. 
Such neural networks have important advantages over neural networks with unconstrained weight
matrices, especially in the case of deep neural networks; in this context, ONNs can
avoid issues with generalization and small gradients \cite{li2019orthogonal}.

Recently, quantum algorithms have been developed which perform the same computation
as ONNs \cite{kerenidis2021classical}. We briefly review the aspects of
this work related to our models in this section.

The quantum ONN (QONN) algorithm of \cite{kerenidis2021classical} takes as input a normalized  $\mathbf{x} \in \mathbf{R}^n$  with $\sum_{i=1}^n x_i^2 = 1$ and determines, to arbitrary precision, the result of applying a matrix $W\in \mathrm{SO}(n)$ acting on $\mathbf{x}$. Put differently, the algorithm rotates vectors on the $n-1$ sphere $S^{n-1}$. 
Because $\mathrm{SO}(n)$ is
a Lie group of dimension $n(n-1)/2$, we require this number
of parameters $\boldsymbol{\theta}$ to specify $W$, and these
parameters are considered to be an input for the quantum algorithm.

The algorithm for computing $W \mathbf{x}$ has three steps which we explain the subsections that follow:
\begin{enumerate}
    \item construct a quantum circuit that loads $\mathbf{x}$ with a unary encoding,
    \item append to the circuit a pyramidal quantum circuit with $n (n-1)/2$ gates parameterized by $\boldsymbol{\theta}$, and 
    \item perform a tomography algorithm to recover the components of $W \mathbf{x}$.
\end{enumerate}

Before explaining these steps further, we describe the two-qubit gate
used throughout as a building block: the reconfigurable beam-splitter (RBS) gate. RBS gates are specified by one parameter $\theta$.
In the computational basis ordered as $\ket{00},\ket{01},\ket{10},\ket{11}$, the RBS gate for a given $\theta$ is
\begin{equation}
\label{eq:rbs-gate}
    RBS(\theta) = 
    \left(
    \begin{array}{cccc}
    1 & 0 & 0 & 0\\
    0 & \cos\theta & \sin\theta & 0\\
    0 & -\sin\theta & \cos\theta & 0\\
    0 & 0 & 0 & 1
    \end{array}
    \right).
\end{equation}

In a quantum circuit diagram, we will use the notation
\begin{center}
\begin{quantikz}
 & \gate[wires=2]{\theta} & \qw \\
 & & \qw
\end{quantikz}
\end{center}
to denote RBS gates with parameter $\theta$.

\subsection{Unary Vector Encoding}
\label{sec:unary}
Consider a vector $\mathbf{x} = (x_1,\ldots, x_n) \in \mathbf{R}^n$.
If $\mathbf{x}$ is normalized, we can consider the quantum state
\begin{equation}
    \label{eq:unary-encoding}
    \ket{\mathbf{x}} = 
    x_1 \ket{100\ldots 00} 
    + x_2 \ket{010\ldots 00} 
    + \ldots
    + x_n \ket{000\ldots 01}.
\end{equation}
This state encodes $\mathbf{x}$ in a simple way.
The encoding is extremely sparse, involving only an
$n$-dimensional subspace of the Hilbert space.

A key property of the RBS gate representing equation \eqref{eq:rbs-gate} is that
they map unary states to unary states. By a unary state, we
mean any state which has the form of equation \eqref{eq:unary-encoding}.
To see this, consider the action of an RBS gate on a unary basis vector $\psi_i = \ket{0\ldots 010 \ldots 0}$. (We are using $\psi_i$ to denote
a computational basis vector which is 0 for all qubits except for the one in position $i$.) If the RBS gate acts on a pair of qubits $a, b$ that
does not include $i$, then it acts as the identity on $\psi_i$. 
On the other hand, if $a=i$ then the RBS gate results in a superposition of $\psi_a$ and $\psi_b$, both of which are unary states.

In fact, we can make a stronger statement. An RBS gate, when acting on any unary
state with the form of equation \eqref{eq:unary-encoding} results in a new unary state
$\ket{x}^\prime$ which is obtained by applying an $\mathrm{SO}(2)$ operation
on the 2-dimensional subspace corresponding to the qubits on which the RBS gate acts.
From this it follows that any combination of RBS gates implements an orthogonal matrix
in the unary encoding.

An explicit construction using only RBS gates for this encoding can be found
in \cite{kerenidis2021classical}. The procedure is to begin
with the state $\ket{100\ldots 00}$ and use
a diagonal stack of $n-2$ RBS gates:
\begin{center}
\begin{quantikz}
 \lstick{$\ket{1}$}& \gate[wires=2]{\alpha_0} & \qw &\qw & \qw \\
 \lstick{$\ket{0}$}& & \gate[wires=2]{\alpha_1} & \qw & \qw \\
 \lstick{$\ket{0}$}& \qw & & \gate[wires=2]{\alpha_2} & \qw \\
  \lstick{$\ket{0}$}& \qw & \qw & & \qw \\
\end{quantikz}
\end{center}
By an $O(n)$ classical computation, we can determine $\alpha_0,\ldots, \alpha_{n-2}$ which construct the desired loaded vector.

\subsection{Pyramidal RBS Construction}
\label{sec:pyramidal-rbs}
In this section, we review the construction of \cite{kerenidis2021classical} to explain how the state
$\ket{\mathbf{x}}$ of equation \eqref{eq:unary-encoding}
can be fed into a simple quantum circuit, built
entirely with RBS gates, to yield the state $\ket{W \mathbf{x}}$ where $W$ is an orthogonal matrix and
the notation, once again, implies a unary encoding of
the resulting vector.

The circuit is a pyramid of RBS gates as follows in the $n=4$ case:
\begin{center}
    \begin{quantikz}
     &\gate[wires=2]{\theta_1}& \qw & \gate[wires=2]{\theta_2} & \qw &\gate[wires=2]{\theta_3} & \qw \\
     &\qw & \gate[wires=2]{\theta_4} &         & \gate[wires=2]{\theta_5} & \qw & \qw \\
     & \qw &   &  \gate[wires=2]{\theta_6} &   & \qw & \qw \\
     & \qw &  \qw &  &  \qw  & \qw & \qw
    \end{quantikz}
\end{center}

As we explained in section \ref{sec:unary}, RBS gates apply orthogonal matrices
to unary states (in the unary encoding). Thus, we are guaranteed that this pyramidal
circuit maps a unary state $\ket{\mathbf{x}}$ to another unary encoded state $\ket{W \mathbf{x}}$ where $W \in \mathrm{O}(n)$. Moreover, the pyramidal
construction applies precisely $n(n-1)/2$ gates which matches the Lie group
dimension of $\mathrm{O}(n)$. This strongly suggests that all matrices $W$ in
$\mathrm{SO}(n)$, the connected component of the identity of $\mathrm{O}(n)$, can be obtained
by appropriately selecting $\boldsymbol{\theta}$.  Construction of parameters
to represent a desired matrix are given in \cite{kerenidis2021classical}.

\subsection{Tomography}
\label{sec:tomography}
With the state $\ket{W \mathbf{x}}$ constructed,
we still need to determine the final output via tomography. \cite{kerenidis2021classical} provides multiple approaches for this. In the experiments of this work, we adopted
one of their techniques which we briefly review here
for convenience. 

Consider a unary-encoded state $\ket{\mathbf{x}}$.
By measuring repeatedly, we obtain $p_1, \ldots, p_n$,
the probabilities of measuring $\psi_1,\ldots, \psi_n$
respectively. These are estimates of $x_1^2,\ldots, x_n^2$ respectively.

The only information that we are missing is signs of $x_1,\ldots, x_n$. It is not possible to determine an overall sign, but if we assume $x_1 > 0$, the following procedure can be used to find all other signs.
We append RBS gates with $\theta=\pi/4$ acting on qubits 1-2, qubits 3-4, 5-6, and so on until $(n-1)$-$n$ if $n$ is even or $(n-2)$-$(n-1)$ otherwise. With this construction, measuring $p_1$ and $p_2$ correspond to $(x_1+x_2)^2$ and
$(x_1 - x_2)^2$. Thus, $p_1 > p_2$ if and only if $x_1$
and $x_2$ have different signs. We can then do the same
construction with the RBS gates offset by one to determine if $x_2$ and $x_3$ have the same signs, and so on.

\section{Vehicle Routing Problems}
This section describes the mathematical
problem to which we apply a hybrid classical-quantum
machine learning model. The problem is 
a very general version of a vehicle routing problem.
This routing problem was described in 
\cite{poc2plus} and, particularly, in \cite{poc3cl}.
I this section, we give a succinct description
of this general vehicle routing problem, and we refer
the reader to \cite{poc3cl} for a detailed explanation.

The goal of the general vehicle routing problem
is to give a problem statement which is very
closely related to an actual supply chain logistics
problem encountered by Aisin Corporation, a Japanese
automotive component manufacturing company. The
routing problem of Aisin Corp. is enormously complicated,
involving many trucks and complex routing requirements
for a large number of boxes.
Our general vehicle routing problem includes the 
salient features of the Aisin Corp. logistics problem
but makes small simplifications to make it possible
to attack the problem with efficient machine learning
techniques.

\subsection{Basic Vehicle Routing Problem}
Vehicle routing problems (VRPs) \cite{dantzig1959truck,toth2002vehicle} come in many forms.
One of the most basic problems starts with the following input:
\begin{itemize}
    \item A collection of locations (nodes) and driving times between all nodes,
    \item one special node called a depot,
    \item a specified amount of \emph{demand} to be picked up from each location and brought to the depot, and
    \item the capacity of a truck (i.e. the amount of demand that a truck can carry).
\end{itemize}
The routing problem then asks to find the minimal driving time path for a truck which starts at the depot and
 fulfills all demand requirements.

\subsection{General Vehicle Routing Problem}
\label{sec:general-vrp}
There are two major features of the logistics problems  of Aisin Corp that are not captured with basic vehicle routing problems: multiple trucks and tensor demand structure.  The presence of multiple trucks is
an obvious challenge that requires little explanation: solutions to multi-truck vehicle routing problems
can involve subtle collaboration among trucks.
Tensor demand structure is a concept described in
\cite{poc2plus, poc3cl} which we now recount and expand upon.

In standard VRPs, there is a special depot node to which
all demand must be brought. However, in a more realistic
problem, requirements are more general. 
For example, we could have the requirement of bringing
demand $d$ from node $2$ to node $3$. Such
a requirement can be stored as a component of a tensor: $D_{2\,3}$
We could also have a requirement like bringing
demand first from node 5 to node 2 and then from
node 2 to node 3.  This can be stored as
a component of a rank-3 tensor $D_{5\,2\,3}$.

Demand tensors can be off-board or on-board.
The examples above are off-board demand because
they describe material which is not on any truck
and is waiting to be picked up. On-board demand
refers to demand currently on a truck. For example,
truck number $m$ can have demand which needs to
be dropped off at node $i$ as its final stop.
We can store this demand as contributing to a tensor $E^m_i$. If truck $m$ has demand which needs to
go to node $i$ but then will need to go to node $j$ (after being picked up again, perhaps by a different truck),
then we can store this as a component to a rank 2 tensor as in $E^m_{ij}$. 

Another situation that arises in realistic situations is
\emph{cyclic} off-board demand. This means, for example,
that demand needs to go from node $i$ to node $j$ and then
back to $i$. This can arise in several logistical setting.
For Aisin Corp., this is a ubiquitous matter: boxes
used to transport parts must be returned to their original departure location. Cyclic demand can also have higher rank: a box may need to go from node $i$ to $j$, then $j$ to $k$, and finally back to $i$. In situations where
we have cyclic demand, we use the term \emph{direct demand} to refer to demand which is not cyclic and we use notation like $D^\mathrm{cyclic}_{ijk}$ and $D^\mathrm{direct}_{ijk}$ to describe cyclic and direct demand tensors. Note that we could describe cyclic demand
with a direct demand tensor of higher rank, but this
is more computationally expensive in practice.

Various types of demands convert between each other
as trucks pick up and drop off material. This can
be understood through the following example adapted from \cite{poc3cl} which gives events
for cyclic initial demand starting at node 3 that must go to node 7 and then node 4 before returning to node 3.
\begin{widetext}

\begin{center}
\begin{tabular}{|c|c|}
\hline 
Component & Description of most recent event.\tabularnewline
\hline 
\hline 
$D_{3\,7\,4}^{\text{cyclic}}$ & Initial material at node 3 waiting to be picked up\tabularnewline
\hline 
$E_{7\,4\,3}^{m=2}$ & Picked up from node 3 by truck 2, next stop node 7\tabularnewline
\hline 
$D_{7\,4\,3}$ & Dropped off at node 7 by truck 2\tabularnewline
\hline 
$E_{4\,3}^{m=1}$ & Picked up from node 7 by truck 1, next stop node 4\tabularnewline
\hline 
$D_{4\,3}$ & Dropped off at node 4 by truck 1\tabularnewline
\hline 
$E_{3}^{m=3}$ & Picked up from node 4 by truck 3, next stop node 3\tabularnewline
\hline 
0 & Dropped off at node 3 by truck 3, requirements fulfilled\tabularnewline
\hline 
\end{tabular}
\end{center}
\end{widetext}

An instance of the general vehicle routing problem is specified by
providing an initial off-board demand tensors $D$, a collection of nodes $z_1,\ldots,z_n$, the number of trucks $N$, a matrix of driving times between nodes $T_{ij}$, and the capacities of trucks $C_1,\ldots,C_N$.

An important construction that we will use in this work is the
concept of \emph{myopic demand vectors}, given by 
\begin{equation}
    \label{eq:onboard-myopic}
    \epsilon^m_i = \sum_{i_2,\ldots, i_r} E^m_{i, i_2,\ldots, i_r}
\end{equation}
and
\begin{equation}
    \label{eq:offboard-myopic-out}
    \delta_i^\mathrm{out} = \sum_{i_2,\ldots, i_r} D^\mathrm{total}_{i, i_2,\ldots, i_r}, 
\end{equation}
where $D^\mathrm{total}$ is the sum of cyclic and direct demand. In words, $\epsilon^m_i$ is the total material on truck $m$ which needs
to be dropped off at node $i$ for its next stop, and $\delta_i^\mathrm{out}$
is the total material at node $i$ that is waiting to be picked up.
We will also define an ingoing myopic off-board demand
\begin{equation}
    \label{eq:offboard-myopic-in}
    \delta_i^\mathrm{in} = \sum_{i_1,i_3,\ldots, i_r} D^\mathrm{total}_{i_1, i, i_3,\ldots, i_r}.
\end{equation}

Of course, we could repeat this construction for any of the
indices, but the further we go to the right in index, the
less ``myopic'' the quantity is.

\subsection{Relationship with Aisin Corp. Logistics}
The general VRP that we described in section \ref{sec:general-vrp}
is inspired directly by the operations of Aisin Corp. in Japan.
They are required, on a daily basis, to ship
340,000 boxes among 21 locations around Aichi Prefecture. These
boxes have either rank-2 or rank-3 cyclic demand requirements:
all boxes must be returned, and some have required intermediate stops
(related to the need to hold at warehouses due to space limitations).

Despite the very large number of boxes, there is substantial redundancy.
There are approximately 15,000 unique boxes, with the 340,000 including
repetitions of identical boxes. Moreover, among the 15,000 boxes,
many have identical routing requirements: there are only 107 unique routing requirements. We refer to the collection of all boxes with a specific routing requirement as a ``box group''. By summing over the volumes of all boxes
within a box group, we can obtain a contribution to the initial cyclic
rank-2 or rank-3 demand tensor. In this way, an instance of the general
VRP can approximate an instance of the realistic logistics problem.

\section{Hybrid Quantum-Classical Attention Mechanisms}
\label{sec:hybrid-policy}
In this section, we present a variation on 
multi-head attention mechanisms that incorporates
the QONN algorithm
described in section \ref{sec:quantum-onn}.
Our model is closely based on the encoder-decoder
of \cite{kool2018attention} as well the generalization
of \cite{poc3cl}. 
In these works, an attention model is used to compute the 
policy of a reinforcement learning agent that learns to
solve combinatorial optimization
problem instances.
We will use our model to solve the general vehicle routing problem discussed in section
\ref{sec:general-vrp}.

\subsection{Standard Attention Mechanism}
\label{sec:standard-attention}
The encoder and decoder of \cite{kool2018attention}
relies on a multi-head attention mechanism which we briefly review in this section. The encoder and decoder of \cite{poc3cl} modify
this structure somewhat, and we review that modification here
as well.

Starting with encoded nodes
$\mathbf{h}_{1}, \ldots, \mathbf{h}_{n}\in\mathbf{R}^{d}$, we compute vectors known as
queries, keys, and values. The queries and keys have a dimension $\alpha$ which is an arbitrary hyperparameter. 
In a single-head attention mechanism, 
we compute a single query, key, and value vector for each encoded node
$\mathbf{h}_{i}$:

\begin{align}
    \mathbf{q}_{i} &= M^\mathrm{query}\,\mathbf{h}_{i} \in\mathbf{R}^{\alpha},\\
    \mathbf{k}_{i}&=M^\mathrm{key}\,\mathbf{h}_{i}\in\mathbf{R}^{\alpha},\\
    \mathbf{v}_{i}&=M^\mathrm{value}\,\mathbf{h}_{i}\in\mathbf{R}^{d}.
\end{align}
Here, each $M$ is a matrix mapping $\mathbf{R}^d$ to either $\mathbf{R}^d$
or $\mathbf{R}^\alpha$.

For the multi-head attention mechanism, we have a positive integer $n_\mathrm{heads}$. For each $s\in \{1, \ldots n_\mathrm{heads}\}$, we have a corresponding query, key, and value map and
thus different queries, keys, and values computed:

\begin{align}
    \label{eq:q-heads-no-source}
    \mathbf{q}_{s\,i} &= M_s^\mathrm{query}\,\mathbf{h}_{i} \in\mathbf{R}^{\alpha},\\
    \label{eq:k-heads-no-source}
    \mathbf{k}_{s\,i}&=M_s^\mathrm{key}\,\mathbf{h}_{i}\in\mathbf{R}^{\alpha},\\
    \label{eq:v-heads-no-source}
    \mathbf{v}_{s\,i}&=M_s^\mathrm{value}\,\mathbf{h}_{i}\in\mathbf{R}^{\beta}.
\end{align}
All of the linear maps $M$ are learned during training. 

An important feature of this mechanism is that the node index $i$ behaves as batch index: the same linear maps are applied to all of the nodes (although different maps are used for different attention heads $s$).
This structure allows sequences to have arbitrary length rather than
hard-coding the sequence length.

An important reason that multi-head attention is used
is to effectively reduce the dimensions of keys, queries, and values. When using a single attention head, we
generally take $\alpha = d$ or at least $\alpha \sim d$.
However, with multi-head attention we can used $\alpha, \beta \sim d / n_\mathrm{heads}$. As we will
discuss in section \ref{sec:quantum-attention},
having a smaller $\alpha, \beta$ makes it possible
to integrate the QONNs
of section \ref{sec:quantum-onn} in a way that
can feasibly be executed on NISQ hardware.

The next step is to compute a \emph{compatibility} for every
query-key pair (for each head). We define

\begin{equation}
\label{eq:basic-compat}
    u_{s\,ia}=\frac{1}{\sqrt{\alpha}} \mathbf{q}_{s\,i}\cdot \mathbf{k}_{s\,a}
\end{equation}
with $\cdot$ denoting a standard dot product. After this, we compute
\begin{equation}
    \label{eq:softmax}
    \rho_{s\,ia}=\frac{\exp(u_{s\,ia})}{\sum_{b}\exp(u_{s\, ib})}\in\mathbf{R}
\end{equation}
which is used to weight a sum over values:

\begin{equation}
    \label{eq:new-nodes-att}
    \mathbf{h}_{s\, i}^{\prime}=\sum_{a}\rho_{s\, ia}v_{s\, a}.
\end{equation}

The only remaining step to to merge the data from the $n_\mathrm{heads}$
attention heads. To do this, we simply concatenate the output from each
head,
\[
     \mathbf{h}_{1\, i}^{\prime} \oplus \ldots \oplus  \mathbf{h}_{n_\mathrm{heads}\, i}^{\prime},
\]
and use a learned linear map with bias on this vector, from dimension $n_\mathrm{heads} \beta$ to dimension $d$ to recover a vector 
$\mathbf{h}^\prime_{i} \in \mathbf{R}^d$.  Here, the symbol $\oplus$
denotes direct sum, which is equivalent to concatenation in this context.

We use the
symbol MHA to denote the function which starts
with encoded vectors $\mathbf{h}_{i} \in \mathbf{R}^d$,
and returns the output sequence
$\mathbf{h}^\prime_{i} \in \mathbf{R}^d$.
MHA computes queries, keys, and values following
equations \eqref{eq:q-heads-no-source}-\eqref{eq:v-heads-no-source}, then computes compatibility and new encoded vectors
for each head, and finally maps to the original encoding dimension.

An important technique used in both \cite{kool2018attention} and  \cite{poc3cl} is the concept of source terms. 
Source terms allow us to modify key, query, and values
in a way that depends on some other quantity, a ``source''. Roughly speaking, the idea is to modify, e.g.,
keys from $\mathbf{k}_i = M \mathbf{h}_i$ to
\begin{equation}
\label{eq:source-concept}
    \mathbf{k}_i = M \mathbf{h}_i + \phi_i \mathbf{u},
\end{equation}
where $\phi_i$ is a real number for each $i$ and $\mathbf{u}$ is a learned vector with dimension $\alpha$, the dimension of keys. The
quantity $\phi_i$ is referred to as a vector source
because it has one index $i$. In general, we
can also have an arbitrary number of sources by summing
over different learned vectors.

As a more concrete of sources, consider the myopic demands $\delta^\mathrm{out}_i, \delta^\mathrm{in}_i$
of equations \eqref{eq:offboard-myopic-out} and \eqref{eq:offboard-myopic-in}. We can use them
to modify equations \eqref{eq:k-heads-no-source} and \eqref{eq:v-heads-no-source}:

\begin{align}
    \label{eq:k-heads-vec-source}
    \mathbf{k}_{s\,i}&=M_s^\mathrm{key}\,\mathbf{h}_{i} + \mathbf{u}_s^\mathrm{key, out}\,\delta_i^\mathrm{out} + \mathbf{u}_s^\mathrm{key, in}\,\delta_i^\mathrm{in},\\
    \label{eq:v-heads-vec-source}
    \mathbf{v}_{s\,i}&=M_s^\mathrm{value}\,\mathbf{h}_{i} + \mathbf{u}_s^\mathrm{val, out}\,\delta_i^\mathrm{out} + \mathbf{u}_s^\mathrm{val, in}\,\delta_i^\mathrm{in}.
\end{align}
Here, $s$ is an index for attention heads, and various $\mathbf{u}_s$
are learned vectors with the same dimension as the left hand side of the equations they appear in; for example, $\mathbf{u}_s^\mathrm{key, out}$
is a vector with the same dimension as keys which is $\alpha$ as specified in equation \eqref{eq:k-heads-no-source}. Note that in this example,
two sources are used for both keys and values.

\subsection{Attention With Quantum Layers}
\label{sec:quantum-attention}
There are a variety of ways that quantum
neural circuits can play a role in attention
mechanisms. 
One approach, developed in \cite{sanches2021short}, 
is to associate a quantum
state for each key and query, and to then apply a unitary operation for each key-query pair and to perform a measurement to obtain a quantum version of key-query
compatibility, replacing equation \eqref{eq:basic-compat}.

In this work, we consider instead a QONN construction. Consider keys,
queries, and values obtained through equation
\eqref{eq:k-heads-no-source}-\eqref{eq:v-heads-no-source}.
For definiteness,
we focus on keys $\mathbf{k}_{s\, i}$ which we take to have dimension $\alpha$ for each head $s$ and each node index $i$.

Now we can apply the QONN
algorithm of section \ref{sec:quantum-onn} to $\mathbf{k}_{s\, i}$. The particular ONN construction
relies on parameters $\mathbf{\theta}$ which specify
the quantum circuit. We will use different parameters for
different attention heads $s$, but, as usual, the
parameters do not depend on $i$. Let $L_s$ denote
the operator which takes as input $\mathbf{k}_{s\, i}$ (for fixed $s$ and $i$), applies the algorithm
described in section \ref{sec:quantum-onn}, including
loading and tomography, and outputs the resulting
classical vector found through tomography. Note
that $L_s$ constructs an $\alpha$-qubit quantum
circuit.

The usage of quantum circuits in individual attention
heads is very well-suited for NISQ quantum hardware. As discussed above, when we use $n_\mathrm{heads}$ attention heads, it is most natural
to use
\begin{equation}
    \alpha, \beta \sim \frac{d}{n_\mathrm{heads}}.
\end{equation}
The circuit construction uses $\alpha$ (or $\beta$ for values) qubits, and the reduction in the number of qubits
allows for implementation on hardware with limited size
while not sacrificing the encoding dimension $d$.
For example, with $d = 128$ and 16 attention heads,
we require only 8 qubit circuits.

Additionally, in a model with $n_\mathrm{heads}$
attention heads, we have a the versatility 
of being able to select only a subset of heads
in which to include QONNs. Thus, we can have non-negative integers $n_\mathrm{classical}$ and $n_\mathrm{quantum}$
with $n_\mathrm{classical} + n_\mathrm{quantum} = n_\mathrm{heads}$. In general, we have the freedom
of choosing these integers differently for
every key, query, and value in all attention heads
throughout the encoder or decoder.

Tuning the number of quantum heads in various 
attention layers is of critical importance
when attempting experiments involving near-term
quantum hardware. The reason for this is not
only limitations due to noisy hardware.
Another major reason relates to access availability
and monetary cost. When submitting jobs to
quantum hardware, time delays can occur due to
jobs queuing. The more separate quantum circuits that need
to be prepared in a workflow, the more serious this
consideration is.  To execute an instance of
the general VRP with a trained encoder-decoder model
including quantum attention heads, the number of quantum
circuits that must be prepared and measured can be
very large.

\subsection{Incorporating Tensor Structure}
\label{sec:dynamical-masking}

We now turn to an element of our attention mechanism
which is unrelated to quantum algorithms.
A major challenge for attention mechanisms is the incorporation
of tensor structure like the tensor demand structure
discussed in section \ref{sec:general-vrp}. The issue can be 
seen through equation \eqref{eq:source-concept}: we can convey 
to an attention mechanism vector quantities which are 
associated with a single
node ($\phi_i$ is associated with node $i$ in 
equation \eqref{eq:source-concept}), but we cannot convey tensor 
quantities that are associated with multiple nodes.

In \cite{poc3cl}, two approaches for incorporating tensor demand
structure were provided. The most powerful method is a \emph{tensor
attention mechanism} where a single query $\mathbf{q}_i$ is used
to probe a sequence of nodes, like $(j,k,l)$. From the sequence
$(j,k,l)$, we obtain a key tensor $K_{jkl}$, which is, for each $(j,k,l)$,  a vector with dimension $\alpha$, so we can compute a compatibility
between node $i$ and nodes $(j, k, l)$. By also constructing
value tensors, we essentially follow equations \eqref{eq:k-heads-no-source}
through \eqref{eq:new-nodes-att}.  The power of the tensor 
attention mechanism is that it allows for tensor sources.
However, to construct keys and value for sequences of tensors,
we are forced to map not from $\mathbf{h}_i$ but instead from
$\mathbf{h}_i \oplus \mathbf{h}_j \oplus \mathbf{h}_k$, incurring
a substantial memory cost in practice.

While we maintain that a tensor attention mechanism is 
the natural approach when dealing with combinatorial
optimization problems with tensor structure,
we can often get away with only including myopic data,
such as equations \eqref{eq:offboard-myopic-out}-\eqref{eq:offboard-myopic-in}. One step further is
to consider matrix myopic off-board demand. In the
case where we have rank 2 and rank 3 cyclic and direct off-board demand, this means
\begin{equation}
    D_{ij} = D^\mathrm{total}_{ij} + \sum_{k}D^\mathrm{total}_{ijk},
\end{equation}
where $D^\mathrm{total}$ is the sum of cyclic and direct off-board demand.

We can incorporate this matrix into the
attention mechanism by using it to modify
compatibility. We replace the dot product with
\begin{equation}
    \label{eq:dynamical-masking-simple}
    \frac{1}{\sqrt{\alpha}} G_{ij} \, \mathbf{q}_i \cdot \mathbf{k}_j,
\end{equation}
where $G$ is some tensor determined by $D$.
This approach can exaggerate query-key compatibility
in cases where $D_{ij}$ is large and suppress compatibility
when the demand is small. 

There are a few reasonable choices for $G$. The
first is $G_{ij} = 1$, which reduces to a basic dot product compatibility.
Next is $G_{ij} = M_{ij}$ where $M$ is the mask defined as $M_{ij} = 1$ when
$D_{ij} > 0$ and $M_{ij} = -\infty$ otherwise. Both of these
are within the methodology of \cite{kool2018attention}. A third
and more novel choice of $G$ is $G_{ij} = \log D_{ij}$. This last
form has several virtues: it reduces to a mask in the sense
that it approaches $-\infty$ as $D{ij}\to 0+$. Moreover, it 
can exaggerate compatibility when $D_{ij}$ is large. A simple
additional adjustment is to use 
\begin{equation*}
    G_{ij} = A D_{ij} + B \log D_{ij} ,
\end{equation*}
which is more sensitive to changes in $D_{ij}$ for larger values.

Rather than having to pick from these various choices,
we can choose all of them
by taking advantage of the multiple heads. In other words,
for a given head $s \in \{1, \ldots, n_\mathrm{heads}\}$, we can put

\begin{equation}
    \label{eq:dynamcal-masking}
    G^s_{ij} = A^s_\mathrm{basic} + A^s_\mathrm{mask} M_{ij} + A^s_\mathrm{log} \log D_{ij} + A^s_\mathrm{lin} D_{ij}.
\end{equation}

\section{Reinforcement Learning and the General VRP}
\label{sec:rl-main}
We now describe how the ideas of the previous sections
can be put together to construct a reinforcement learning
policy.

Reinforcement learning concerns Markov decision processes
where an agent performs an action when presented
with the state of its environment and, as a stochastic 
result of the action, the state changes and the agent
receives a reward. The action performed by
the agent is determined by the agent's policy
which is a function $\pi$ that, given the state
$s$, gives the probability $\pi(a\,|\,s)$ of performing a given action $a$ 
.

In the context of combinatorial routing problems like
the traveling salesman problem, the action of an
agent would be the choice of which node to travel
to next and the environment state would be the
prior route. For a vehicle routing problem
with a single truck, the situation is quite similar
except that we also need to include in the state
all prior information about the current demand structure:
\[
\pi\left(z\,|\,(\xi_0, \ldots, \xi_{t-1}), D_{t-1}\right),
\]
where $\xi$ denotes the prior route and $z$ is a proposed node to drive to next. Note that we are using $D_{t-1}$ loosely to include all relevant information about demand.

For problems like this, the probability of choosing
an entire route $\xi$ is a product of probabilities
at each step as long as the policy is fixed
during the episode
\begin{equation}
\label{eq:total-route-policy}
    \pi(\xi \, | \, D_0) = \prod_{t=1}^k \pi\left(\xi_t \,|\,(\xi_0, \ldots, \xi_{t-1}), D_{t-1}\right).
\end{equation}
This structure is important because we will use
the REINFORCE algorithm \cite{reinforce-williams}
which depends on the logarithm of the probability
of choosing an action. The multiplication in equation
\eqref{eq:total-route-policy} converts to a summation,
allowing us to deal with the probability of the
entire route without it being too small to work with in practice.

In addition, the general VRP expands beyond a policy for a single truck and must account for other trucks as part of the environment, thus enabling the policy benefit from cooperation between multiple trucks.  We have implemented such a general VRP model by devising a scheme that applies to teams of trucks in which at any given moment in time one truck is the \emph{active} truck and the other \emph{passive} trucks are part of the environment.  The training proceeds through epochs of time in which the effects of the other trucks come into play.  The overarching approach is described below.

\subsection{Encoder-Decoder Policy}
\label{sec:encoder-decoder-policy}

The machine learning model that we use as a model
for $\pi$ is an encoder-decoder attention model closely based on \cite{kool2018attention} and \cite{poc3cl}. In fact,
we use almost the exact model of \cite{poc3cl} except
for the role of quantum circuits. We therefore
refer the reader to \cite{poc3cl} for the details
of the encoder and decoder layers and only
highlight the hyperparameters and the use of QONNs here.

The basic idea of the policy model is that
the nodes $z_1, \ldots, z_n$ along with
information about the initial demand can be thought of
as a sequence of data. We start with coordinates of the
nodes, $\mathbf{x}_i \in \mathbf{R}^2$, which we take
to lie in the unit square. We append information
about initial demand structure
\begin{equation}
    \overline{\mathbf{x}}_{i}
    =\mathbf{x}_{i} \oplus 
    \left(\delta_{i}^{\text{in,init}},\delta_{i}^{\text{out,init}}\right)
\end{equation}
and use the sequence $\left(\overline{\mathbf{x}}_1, \ldots \overline{\mathbf{x}}_n\right)$ as the input to an encoder.
The encoder treats $i$, the node index, as a batch dimension.

Encoding proceeds through layers, and each encoder layer contains an attention sub-layer
(equation \eqref{eq:enc-layer-att})
followed by a feed-forward sub-layer (equation \eqref{eq:enc-layer-ff}):

\begin{equation}
\label{eq:enc-layer-att}
    \tilde{\mathbf{h}}^{l-1}_{i} = 
    \mathrm{BN}\left(
    \mathbf{h}^{l-1}_{i}  + \mathrm{MHA}\left(\mathbf{h}^{l-1}_{i}\right) 
    \right),
\end{equation}

\begin{equation}
\label{eq:enc-layer-ff}
    \mathbf{h}^{l}_{i} =
    \mathrm{BN}\left(\tilde{\mathbf{h}}^{l-1}_{i}  + \mathrm{FF}\left(\tilde{\mathbf{h}}^{l-1}_{i}\right) \right).
\end{equation}

In these equations, $\mathrm{BN}$ is a batch normalization layer
\cite{ioffe2015batch}, $\mathrm{FF}$ is a feed-forward
network, and MHA is a multi-head attention 
layer as described in section \ref{sec:standard-attention}
along with dynamical masking 
from \ref{sec:dynamical-masking}. 
The feed-forward
layers have a single hidden dimension $d_\mathrm{ff}$ and
consist of a linear layer with bias mapping $\mathbf{R}^d \to \mathbf{R}^{d_\mathrm{ff}}$ followed by a ReLU activation function,
a dropout layer, and finally a linear map with bias back to $\mathbf{R}^d$.

The output of the final encoder layer gives encoded nodes
$\mathbf{h}_i \in \mathbf{R}^d$. These nodes,
along with contextual information about the current
environment state, are then decoded. Decoding
yields a probability distribution over possible
nodes for the active truck to select as its next
destination.

Our encoder and decoder structure are identical
to that of \cite{poc3cl} other than the usage
of quantum circuits. The encoder consists of three attention layers.
As in \cite{poc3cl}, we use dynamical masking 
(equation \eqref{eq:dynamcal-masking}) rather
than the tensor attention mechanism due to
memory limitations. 

Regarding the inclusion of quantum circuits,
we use two different models:
\begin{enumerate}
    \item \emph{Simulation-only model}: A quantum orthogonal neural network is used for every key, query, and value in the encoder and decoder including all three encoding layers. We use $d=128$ for encoding and 8 attention heads. There are 16 qubits for circuits with $\alpha = d/ 8$.
    \item \emph{Hardware experiment model}: We only
    use a quantum orthogonal neural network for queries
    and keys in the encoder. We do not use quantum
    circuits for the decoder. We use $d=64$ and eight attention heads with 8 qubits per head.
\end{enumerate}

\subsection{Training and Simulation}
For the purposes of training, our
model was written entirely in PyTorch.
In particular, quantum circuits are also emulated
through PyTorch. The quantum circuit emulation
approach is direct state vector manipulation:
we load a unary state by hand, stored as
a PyTorch \texttt{Tensor}, and we apply
gates through matrix multiplication on appropriate
subsystems. The final readout is also done
in an nonphysical fashion: we simply read
the unary components final state vector.

This approach has a major advantage of being easy
to run efficiently for fewer than $\sim 30$ qubits (beyond which point, the memory cost of storing state vectors becomes
problematic). One disadvantage, however,
is that we do not account for stochastic behavior
of quantum computing, either due to error or simply because
of the probabilistic aspect of measurement.

\subsection{Learning Algorithm}

Following \cite{kool2018attention} and \cite{Sutton1998}, we implemented a variant of REINFORCE \cite{reinforce-williams} but now adapted for quantum circuits.  Our REINFORCE variant is given in algorithm  \ref{alg:reinforce-variant} below and is more fully described in \cite{poc3cl}. Note that this algorithm is broken up into epochs and batches.

For our purposes we take a variant of REINFORCE adapted to our vehicle routing problem as follows. First,
for the purposes of the algorithm, we take the entire episode to be defined
by a single action. In other words, the episode is just $a_0, r_1$. The action
$a_0$ is the entire route specification $a_0 = (\xi_1, \xi_2, \ldots \xi_k)$ where each $\xi_i$ are nodes. The reward $r_1$ is simply the negation of the route length (or time) $-L(\xi)$ which is computed by summing the appropriate distances between nodes based on a metric or on known travel times.

The second important aspect of our variant is the baseline methodology. This idea is roughly adapted directly from \cite{kool2018attention}. This essentially means that some function $b$ of states (but not of actions) is constructed with each episode and the return $G$ in the algorithm is replaced by $G - b(s)$. This algorithm still converges to the optimal policy theoretically and, with a well-chosen baseline, does so much faster.  In this work, we maintain a ``baseline agent'' which uses the same parameterized policy but does constantly update its parameter $\theta$. Instead, the baseline agent uses an outdated parameter $\theta_{\text{BL}}$ which is occasionally updated to match the primary agent's $\theta$, but only when the agent substantially and consistently outperforms the baseline.

\begin{algorithm}
\caption{REINFORCE variant for VRP}\label{alg:reinforce-variant}
\begin{algorithmic}
\STATE Input: Parameterized policy $\pi$
\STATE Input: Integers \texttt{num\_epochs, batch\_size, batches\_per\_epoch}
\STATE Input: Initial parameter $\theta$
\STATE $\theta_{\text{BL}} \gets \theta$
\FOR {$e =1, \ldots,$ \texttt{num\_epochs}}
    \FOR{$b =1, \ldots,$ \texttt{batches\_per\_epoch}}
        \STATE $\xi \gets$  (\texttt{batch\_size} many episodes from $\pi(\theta)$)
        \STATE $\xi_\text{BL} \gets$  (\texttt{batch\_size} many episodes from $\pi(\theta_\text{BL})$)
        \STATE $\nabla J \gets \texttt{batch\_mean}\left( 
            (L(\xi) - L(\xi_\text{BL}) \nabla_\theta \log \left(\sum_{i=1}^k \pi(\xi^i, \theta) \right)
        \right)$
        \STATE $\theta \gets \textrm{descent}(\theta, \nabla J(\theta))$
    \ENDFOR
    \IF {\texttt{baseline\_test()}}
        \STATE $\theta_\text{BL} \gets \theta$
    \ENDIF
\ENDFOR
\end{algorithmic}
\end{algorithm}

The summation $\sum_{i=1}^k \pi(\xi^i, \theta)$ is a sum over the probabilities computed by the encoder/decoder network at each stage of the route. $k$ refers to the number of steps in the route and the index $i$ runs over steps in the route, not over batch entries. The entire computation is performed for each batch entry and averaged over.

\section{Supply Chain Management Workflow}
While the experimental approach of section
\ref{sec:encoder-decoder-policy} is restricted to a modest system size,
we can use the simulation-only model of section 
\ref{sec:encoder-decoder-policy} to test our
model at a more ambitious scale. In fact,
through a decomposition technique of breaking
the full supply chain problem into subsets of a few
trucks at a time, we can deploy our model on 
a full-scale problem to meet the demands of
a commercial supply chain.

We thus turn to the problem of finding solutions to the
Aisin Corp. routing problem instance (which we refer to as the AVRP).

\subsection{Node Subset Search}
\label{sec:node-subset}
Consider an instance of the general VRP with $n$ nodes
 and demand structure $D^\mathrm{init}$. Given the demand structure,
the time matrix, and the driving window $T_\mathrm{max}$,
we can estimate the number of trucks $N$ that will be necessary to fulfill all requirements.

Suppose that we have an algorithm to find solutions to general
VRPs with a smaller number of nodes and trucks and with
smaller demand structure. Our algorithm works for $n^\prime < n$ nodes and
$N^\prime < N$ trucks.

This situation arises in our context naturally: we can train, for instance,
an agent to solve general VRP instances with 10 nodes and three trucks,
a scale smaller than the AVRP with its 21 nodes and over 100 trucks.

We should be able to use our smaller-scale algorithm to solve the larger
problem by applying it repeatedly to different subsets to nodes to gradually
fulfill all demand requirements. This raises a question of how
to find good subsets. 

We begin by looking at the demand structure $D^\mathrm{init}$. For simplicity, assume that this consists only of rank-3 cyclic demand 
(other cases are very similar and we describe the necessarily modifications below). From $D^\mathrm{init}$ we can
identify the nonzero components. These are tuples of nodes
\begin{align*}
    i_1, &j_1, k_1\\
    i_2, &j_2, k_2\\
    &\vdots\\
    i_u, &j_u, k_u\\
\end{align*}
where $u$ is some integer (which happens to be 107 for the AVRP, corresponding to the 107 unique routing requirements). We can
begin our node search by uniformly randomly selecting one of these triples of nodes from the list. (In cases where demand also includes rank 2 or another
rank, we include tuples of appropriate length in the list for nonzero demand
cases and we allow such tuples to be selected as well.) After drawing 
a tuple from the list, we remove it from the list.
Suppose that we select the tuple $(i_3, j_3, k_3)$. 
We define a starting subset of nodes as $A = \{i_3, j_3, k_3\}$.
If $3 < n^\prime$,
we continue to draw nodes. Suppose that we next draw $(i_5, j_5, k_5)$.
We now consider the set $A=\{i_3, j_3, k_3, i_5, j_5, k_5\}$. If any
node repeats (for instance, if $i_3 = j_5$), that entry is only counted once in the set. There will now be between 4 and 6 elements in $A$.
If $|A| < n^\prime$, we continue and otherwise we stop. Continuing in this
way, we can either eventually run out of tuples or we can reach $|A| \geq n^\prime$. If we reach $|A| = n^\prime$, we stop and use $A$ as our
first guess of a node subset. If $|A|$ exceeds $n^\prime$, we remove
the most recently added subset. If we run out of tuples, we stop.

After this process, we might have $|A| < n^\prime$. In this case,
we simply randomly add additional nodes outside of $|A|$ until reaching 
$|A| = n^\prime$.

At this point, we have obtained a random node subset $A$. We repeat this
process $K$ times to obtain $k_\text{node draws}$ different random subsets, and we apply our algorithm $k_\text{subset attempts}$ times
for each of the $k_\text{node draws}$ subsets. We compute
the mean demand fulfilled for the $k_\text{subset attempts}$ trials,
and we select the node subset with the highest mean demand coverage.

\subsection{Execution Loop}
\label{sec:execution-loop}
With a technique for selecting node subsets, we are now in a position
to describe the execution procedure. This begins with the initial
demand structure $D^\mathrm{init}$ which is determined by the AVRP's
requirements. Next, a node subset is selected through the node
search technique of section \ref{sec:node-subset}. We identify
the portion of $D^\mathrm{init}$ that is supported by the subset
and we map it onto a demand structure $D^\prime$ for $n^\prime$ nodes.
To regulate the policy input, we clip $D^\prime$ at some maximum value $C$.
We then apply the trained agent $k_\text{execution trials}$ times
and select the trial with the highest percentage of covered demand.
This trial has a specific routing for $N^\prime$ trucks and
corresponds to a certain demand fulfillment. The route as well as
the on-board and off-board demand at each step is saved and
the initial demand $D^\mathrm{init}$ is modified: it is reduced
by the amount of demand satisfied by the route. 

This process is repeated, each time first performing node selection
and then finding the best route out of $k_\text{execution trials}$ attempts.
We repeat until all demand is satisfied. The total number of iterations
of this procedure multiplied by $N^\prime$ will be the total number
of trucks needed for our solution.

\subsection{Simulated Execution}

The result of the execution loop yields approximate solutions to 
the general VRP. We obtain a collection of routes
for various trucks $x_{m\, \tau}$ where $m$ ranges over trucks
and $\tau$ over time steps for each truck.
However, we still need a way to convert such routes to candidate
solutions to the overall AVRP.

 Our strategy is to interpret the routes $x$ as ``suggested routes'' and to attempt to use them in a \emph{full-scale} supply chain simulation. We make use  of the simulation of \cite{poc3cl}. Every box is individually
 tracked, has rank-2 or rank-3 requirements, and
 we require that demand is cyclic--boxes must be returned to their origin node.
 Each box has a specific individual volume. The simulation is designed to be
 as similar as possible to the actual commercial routing problem of Aisin Corporation.

 Trucks follow along the routes $x_{m,\tau}$ that they are assigned from the execution algorithm and they pick up boxes as is appropriate for their route. Boxes a picked
 up according to the algorithm described in \cite{poc3cl}, and we carefully
 ensure that trucks only drive within the allowed drive time windows (two 8-hour shifts).
 The result of full-scale simulation is that, rather than having a list of suggested
truck routes, we obtain a precise statement about what each truck in the supply chain
is doing at all times, including exactly which boxes must be picked up and dropped off
at various nodes.

\section{Results: Experiment With Ion Trap Quantum Hardware}
\label{sec:ionq-exp}

In \cite{poc3cl} we implemented the entire supply chain management workflow classically using standard machine learning algorithms.  In this work, we implement and test the workflow using quantum circuits of the ONNs implemented on real quantum computing hardware.  First, we return our focus to the discussion of the quantum circuits in section \ref{sec:quantum-onn} and discuss the essential experiments needed to test their performance.  Then we discuss the implementation of the neural network and attentions heads.  And finally we discuss the performance of the quantum circuit in generating solutions when used as part of the overall workflow.

\subsection{Hardware Experiment for QONN Accuracy}
\label{sec:ionq-benchmark}

Current quantum hardware is limited by noise and the precision of the gates they can implement, so there is an important question: how accurate is the QONN algorithm when run on IonQ?  

To answer this, we performed the following experiment using the 11-qubit IonQ device.  The IonQ device was chosen because it has relatively good noise levels and the inherent full-connectivity of the qubits simplifies the embedding of the algorithmic circuit onto the hardware graph, thereby avoiding the need for additional gates operations to implement gate interactions between qubits that are not directly connected physically.  Additionally, the IonQ device is commercially available on Amazon Web Services Braket cloud-based quantum platform, and this makes it available to access by the authors and other researchers via a straightforward subscription service \cite{wright2019}.  Despite the emerging commercial availability to quantum computing hardware, it is still an expensive resource, so users must be careful to tune their algorithms to get the best performance for an affordable cost.

For various number of qubits (4, 5, 6, 7, 8, 9, and 10 qubits), we
constructed the pyramidal RBS circuit of section \ref{sec:pyramidal-rbs}. We used QC Ware's commercially available implementation of unary data loader circuits, and we used the tomography method described
in section \ref{sec:tomography}. We used randomly selected RBS parameters and also randomly
selected input vectors. All parameters and vectors were randomly drawn from a
normal distribution with mean 1 and variance 1. For each fixed number of qubits
(4, 5, 6, 7, 8, 9, and 10), we selected 10 different randomly drawn parameter sets.
We performed 500 measurements on each of the 3 circuits required for unary
tomography. We therefore ran 210 quantum circuits with 500 measurements for
each, a total of 105,000 measurements.

\begin{figure}
\centering
\includegraphics[width=0.6\textwidth]{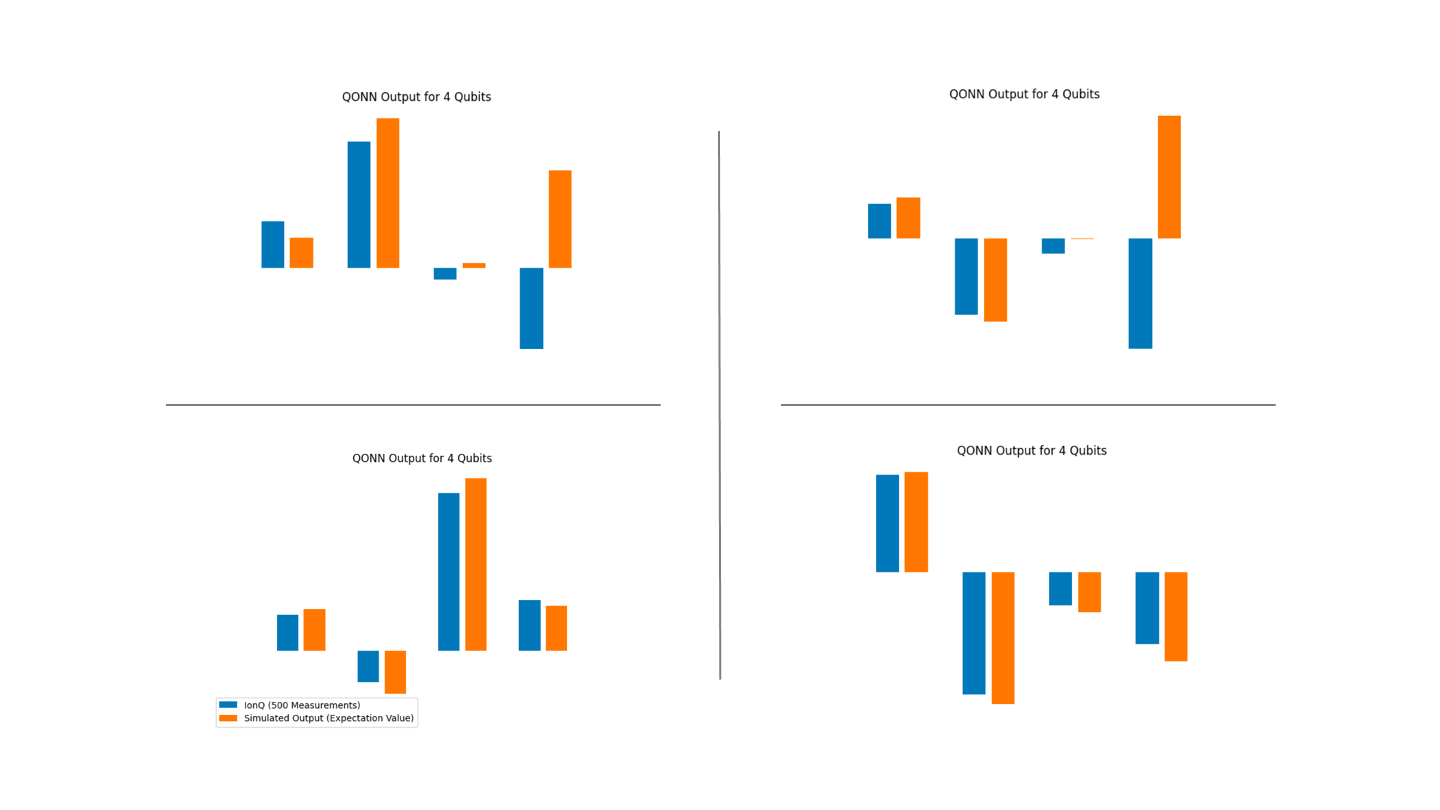}
\caption{Results of 4-qubit QONN tomography}
\label{fig:ionq4}
\end{figure}

\begin{figure}
\centering
\includegraphics[width=0.6\textwidth]{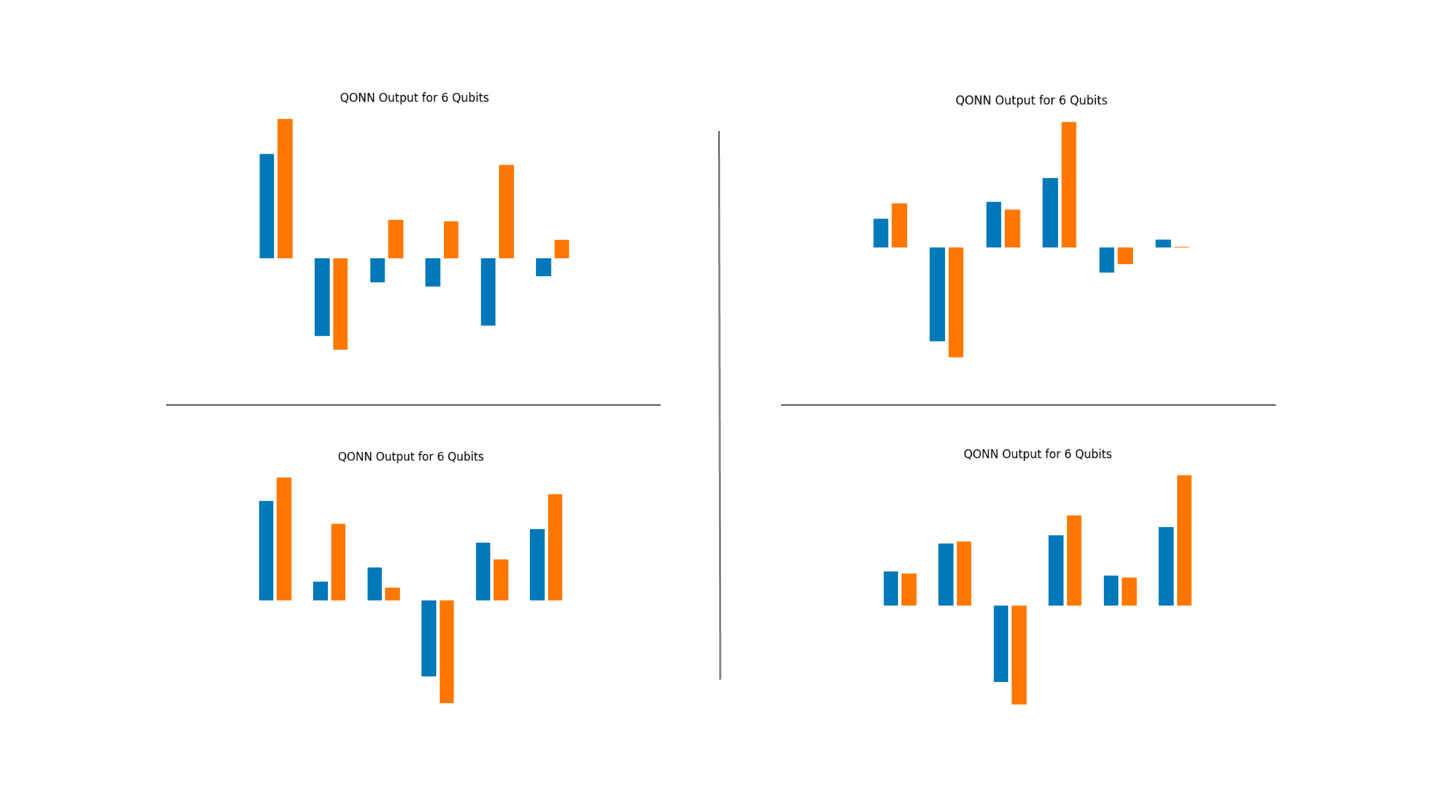}
\caption{Results of 6-qubit QONN tomography}
\label{fig:ionq6}
\end{figure}

\begin{figure}
\centering
\includegraphics[width=0.6\textwidth]{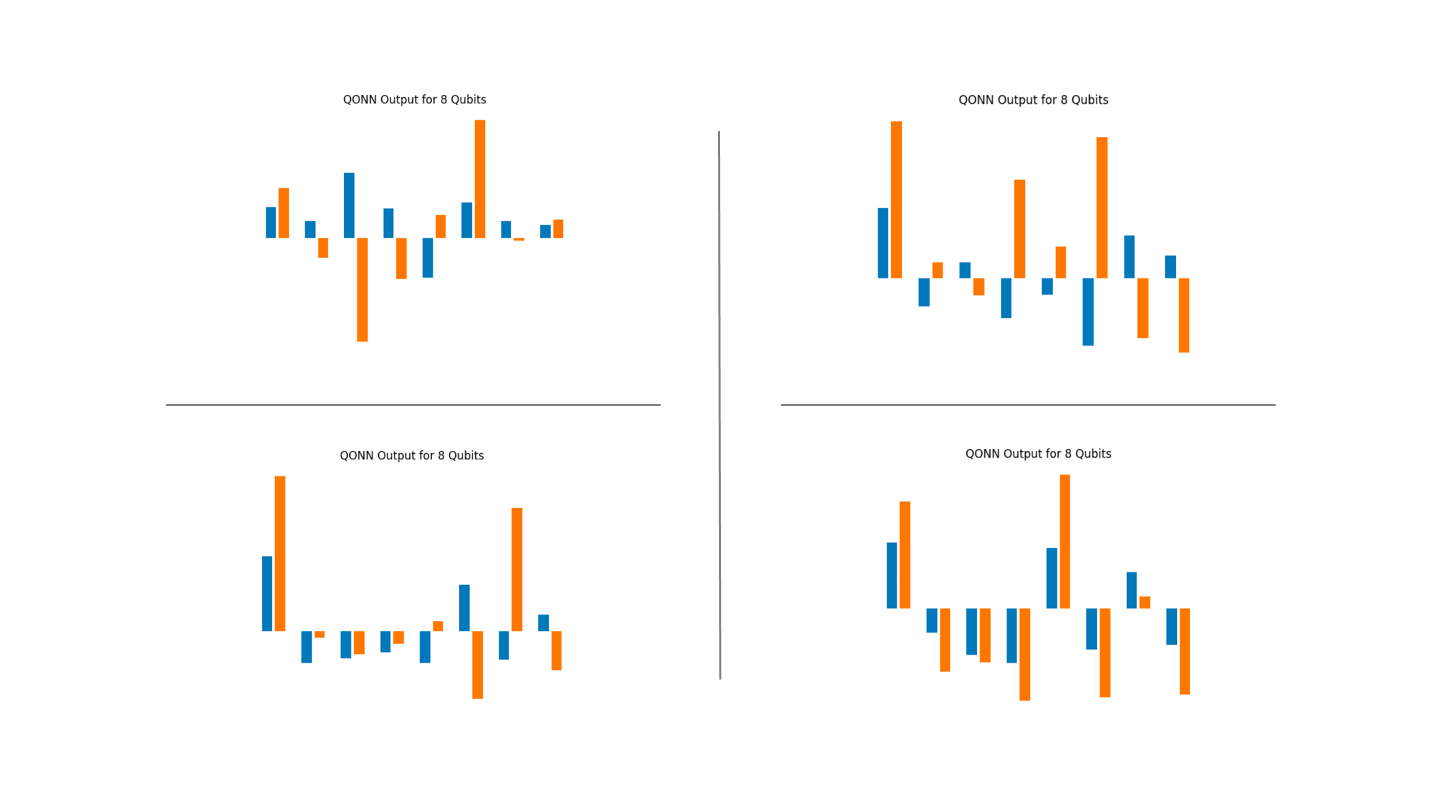}
\caption{Results of 8-qubit QONN tomography}
\label{fig:ionq8}
\end{figure}

\begin{figure}
\centering
\includegraphics[width=0.6\textwidth]{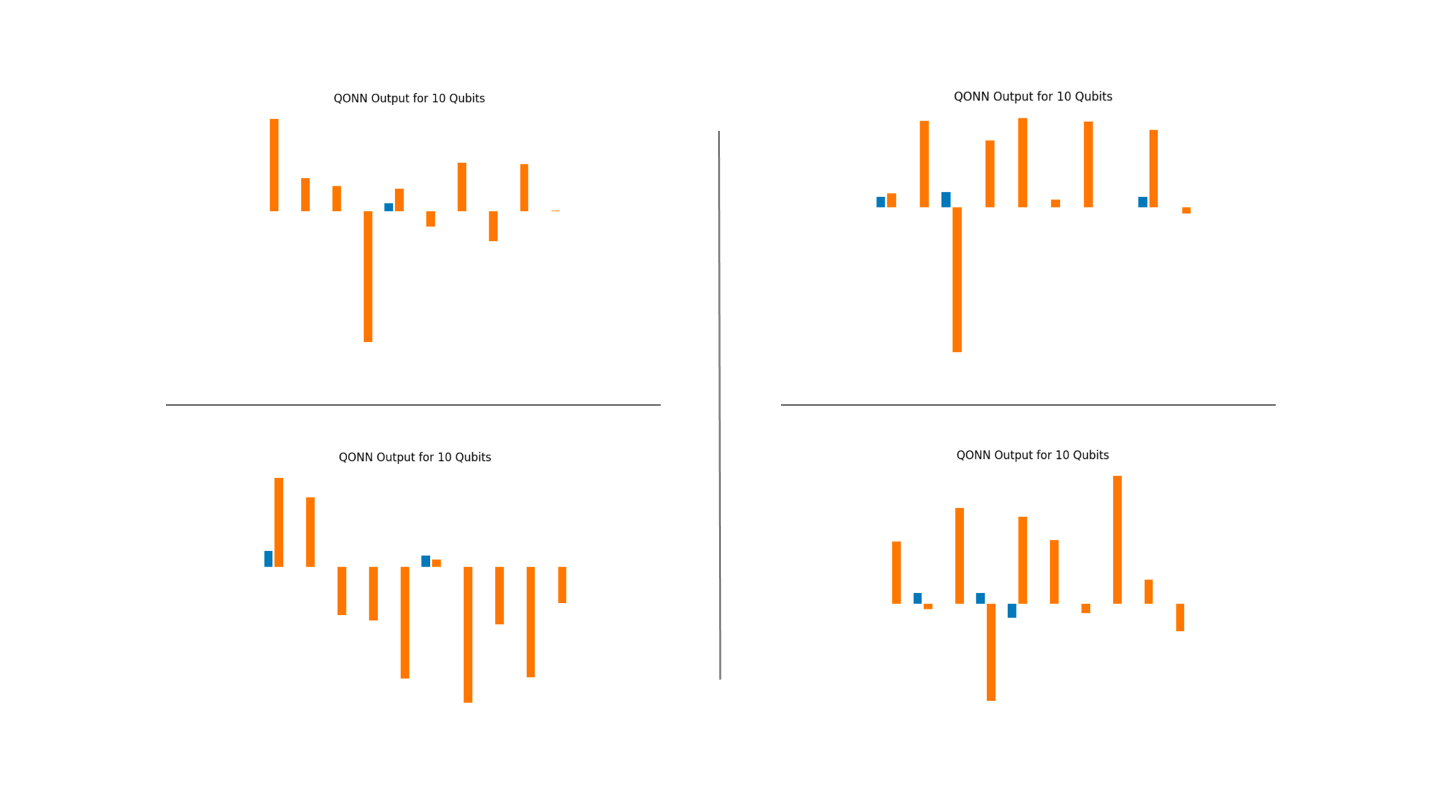}
\caption{Results of 10-qubit QONN tomography}
\label{fig:ionq10}
\end{figure}

A sample of our results are shown in figures \ref{fig:ionq4}-\ref{fig:ionq10} for the cases of 4, 6, 8, and 10 qubits.  In figure displays a montage of four runs selected to show the variety of results returned from each tomography measurement accomplished according to the parameters described in the preceding paragraph.  Since the circuits are random, each of the four tomography results will look different.  However the error is revealed in comparing the hardware (blue) and emulated (orange) pairs for each qubit.  For an $N$-qubit tomography experiment, you will see $N$ pairs of bars.  In each pair we compare the hardware results against the classically computed emulation of the circuit.  Where the hardware results (blue) differ from the emulation results (orange), the hardware results are in error.  

In figure \ref{fig:ionq4}, the 4-qubit case, there is generally good agreement between all pairs with an occasional sign discrepancy.  As we include more qubits in the ONN, one can see evermore degraded performance as the noise and imprecision accumulate to corrupt the result, so that by the 10-qubit case, the hardware results have become almost completely random, averaging to zero.

%\begin{figure}
%\centering
%\includegraphics[width=0.6\textwidth]{ionq_5.png}
%\caption{}
%\label{fig:ionq5}
%\end{figure}

%\begin{figure}
%\centering
%\includegraphics[width=0.6\textwidth]{ionq_7.png}
%\caption{}
%\label{fig:ionq7}
%\end{figure}

%\begin{figure}
%\centering
%\includegraphics[width=0.6\textwidth]{ionq_9.png}
%\caption{}
%\label{fig:ion9}
%\end{figure}

Our insights are as follows:

\begin{itemize}
    \item For up to 6 qubits, the performance of IonQ is quite good.
    \item For up to 8 qubits, the performance of IonQ is sufficient to capture aspects of the desired QONN behavior.
    \item At and beyond 9 qubits, we found that QONNs are not appropriately captured by IonQ with its current capabilities.
    \item Occasionally signs out outputs can be flipped. This is a tomography error. Such errors can cause a cascade of errors: if the wrong sign for qubit 3 is selected, then we get the wrong signs for qubits 4, 5, 6, ... unless another error corrects the original error.
    \item We emphasize that the errors in quantum computing we are seeing here are not surprising and reflect the early state of quantum hardware.   
\end{itemize}

\subsection{Results: AVRP Training with Ion Trap Quantum Hardware}
\label{sec:ionq-workflow}

After testing and diagnosing the maximum number of qubits that could be reliably used in the QONN, we then tested how well the QONN can support a hybrid the entire AVRP workflow.  First we needed to select a problem size suitable for the quantum hardware.  We selected 8 of the most active nodes from actual
Aisin Group data. Specifically, we selected
the following nodes:

\begin{enumerate}
    \item NISHIO CROSS-DOCKING
    \item NO.1 AND 2 PLANT
    \item OKAZAKI AND ELECTRIC PLANT
    \item OKAZAKI EAST PLANT 
    \item TAHARA PLANT 
    \item GAMAGORI PLANT
    \item KIRA PLANT 
    \item MEIKO
\end{enumerate}

These nodes actually have cyclic rank-2 and 3 demand
that flows between them. Due to limitations
of current quantum hardware, we choose to only
consider direct rank-2 demand. We sum over all
demand and use the summation as a contribution
to an effective rank-2 demand. We then drop the
box return constraint. When hardware becomes
more advanced, we should be able to
include more constraints.

We know from the results of section \ref{sec:ionq-benchmark} that at the 8 qubit scale, we are pushing the limits of current hardware for QONN implementation.  For this 8-node problem scenario, we trained a variety of models, classically and quantumly, with slightly different problem parameters to study and compare the performance.  

\begin{figure}
\centering
\includegraphics[width=.5\textwidth]{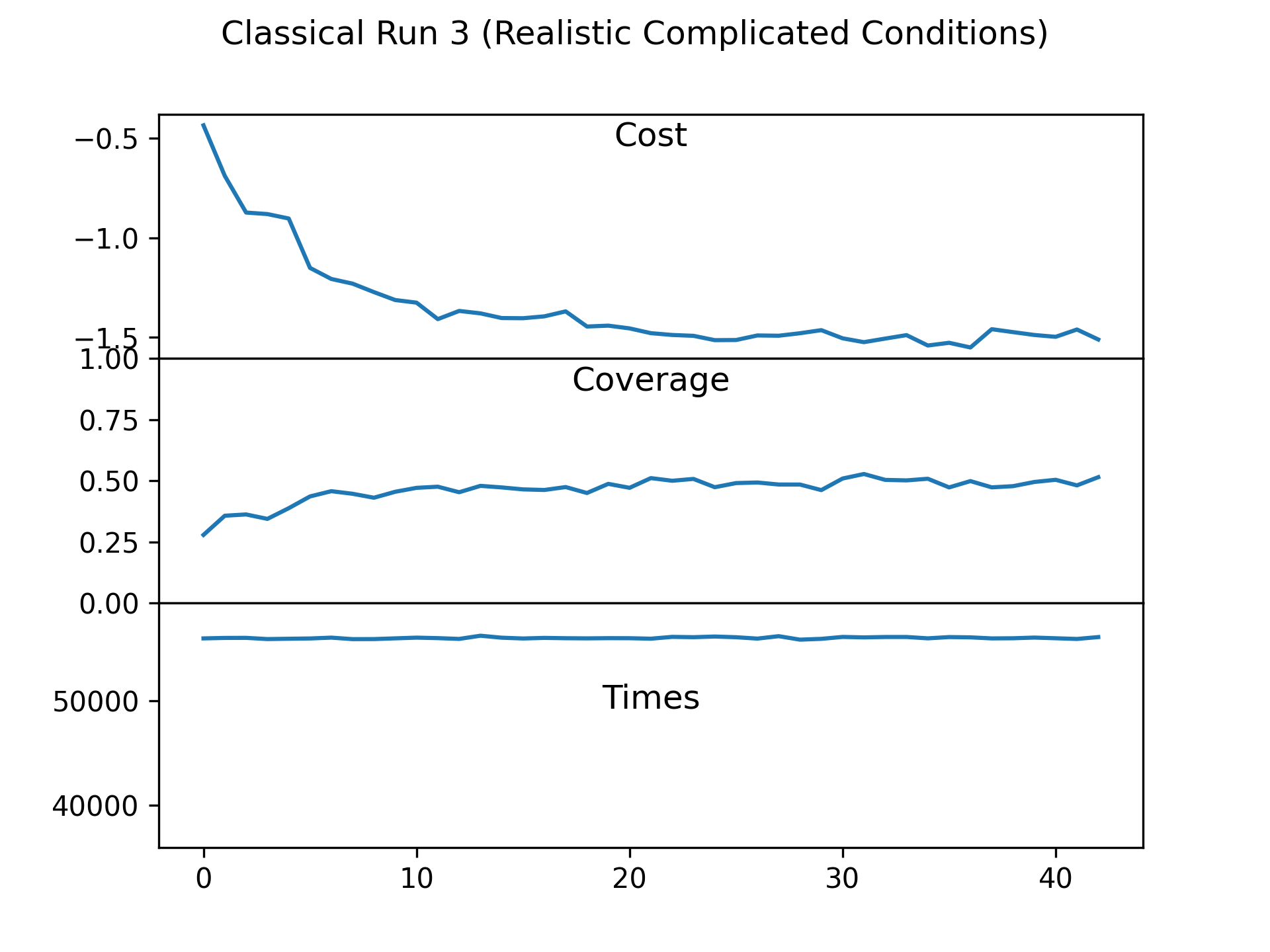}
\caption{Classical training run.  Highly realistic 3 truck, rank 2 and 3 demand, with
cyclic box-return constraint. Uses 8 nodes.}
\label{fig:classical3}
\end{figure}

First we show the result from a classical training run to set a baseline of training performance.  Figure \ref{fig:classical3} shows the classical training results versus training epoch for an 8-node, 3-truck, including rank 2 and rank 3 demand structure, and with cyclic box-return constraint.  

The cost is a dimensionless cost function, the coverage is a normalized variable of the total demand (1.00 = 100\%), and the Times represent how much time in seconds was used to complete the problem (must be less than the 16 hours = 57,600 seconds constraint).

\begin{figure}
\centering
\includegraphics[width=.5\textwidth]{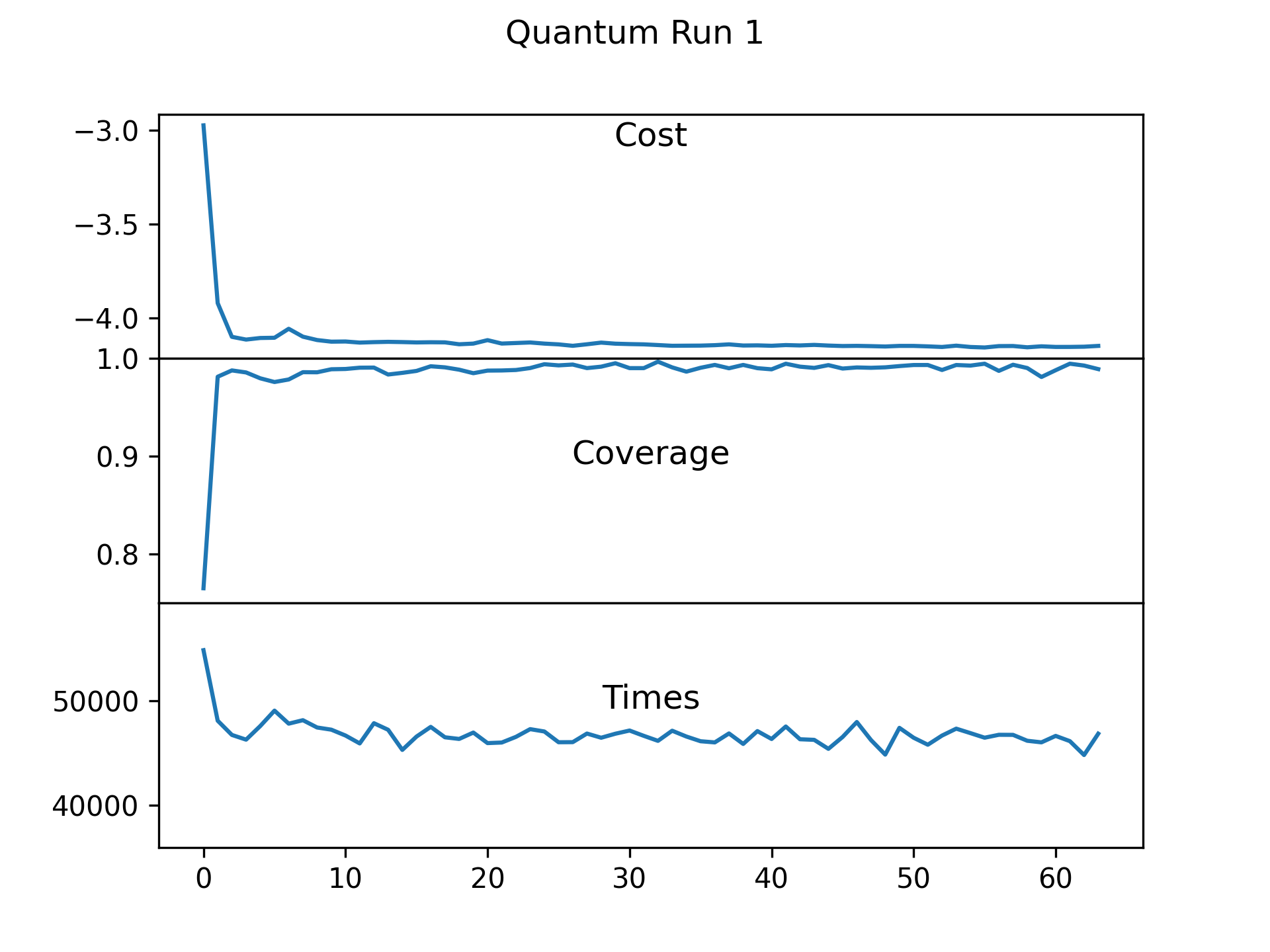}
\caption{Quantum training run.  2 truck, rank 2 demand, with
no cyclic box-return constraint. Uses 8 nodes.}
\label{fig:quantum1}
\end{figure}

Now we contrast that with two quantum circuit training runs.  Figure \ref{fig:quantum1} shows the quantum training results versus training epoch for a simplified 8-node, 2-truck, including rank 2 only demand structure, and with  no cyclic box-return constraint.

\begin{figure}
\centering
\includegraphics[width=.5\textwidth]{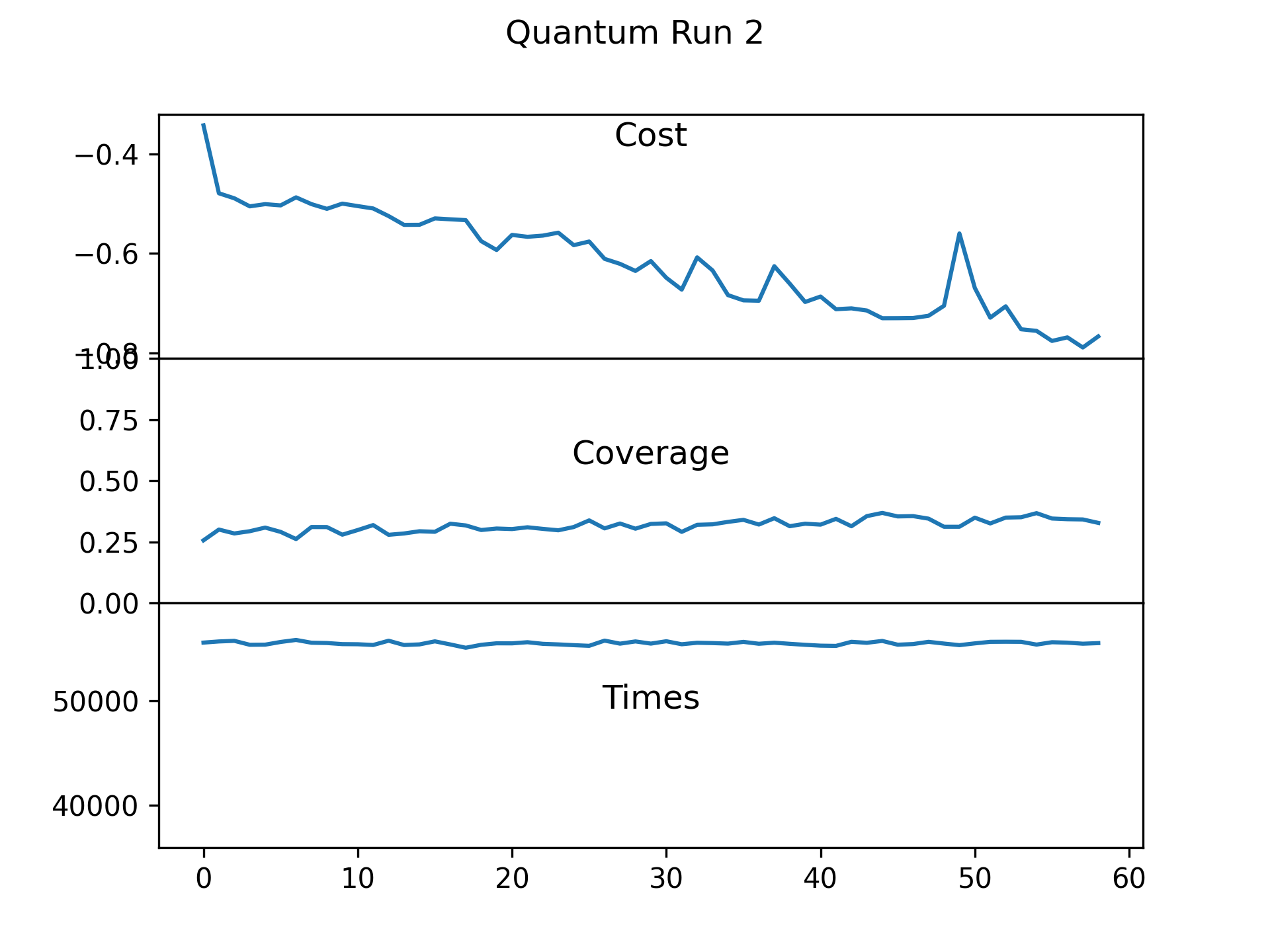}
\caption{Quantum training run.  2 truck, rank 2 and rank 3 demand, with
no cyclic box-return constraint. Uses 8 nodes.}
\label{fig:quantum2}
\end{figure}

Adding the more realistic rank 3 demand structure, however, significantly hinders the training performance.  Figure \ref{fig:quantum2} shows the quantum training results versus training epoch for the same problem as before, but now with the more realistic rank 2 and rank 3 demand structure.

\begin{figure}
\centering
\includegraphics[width=.5\textwidth]{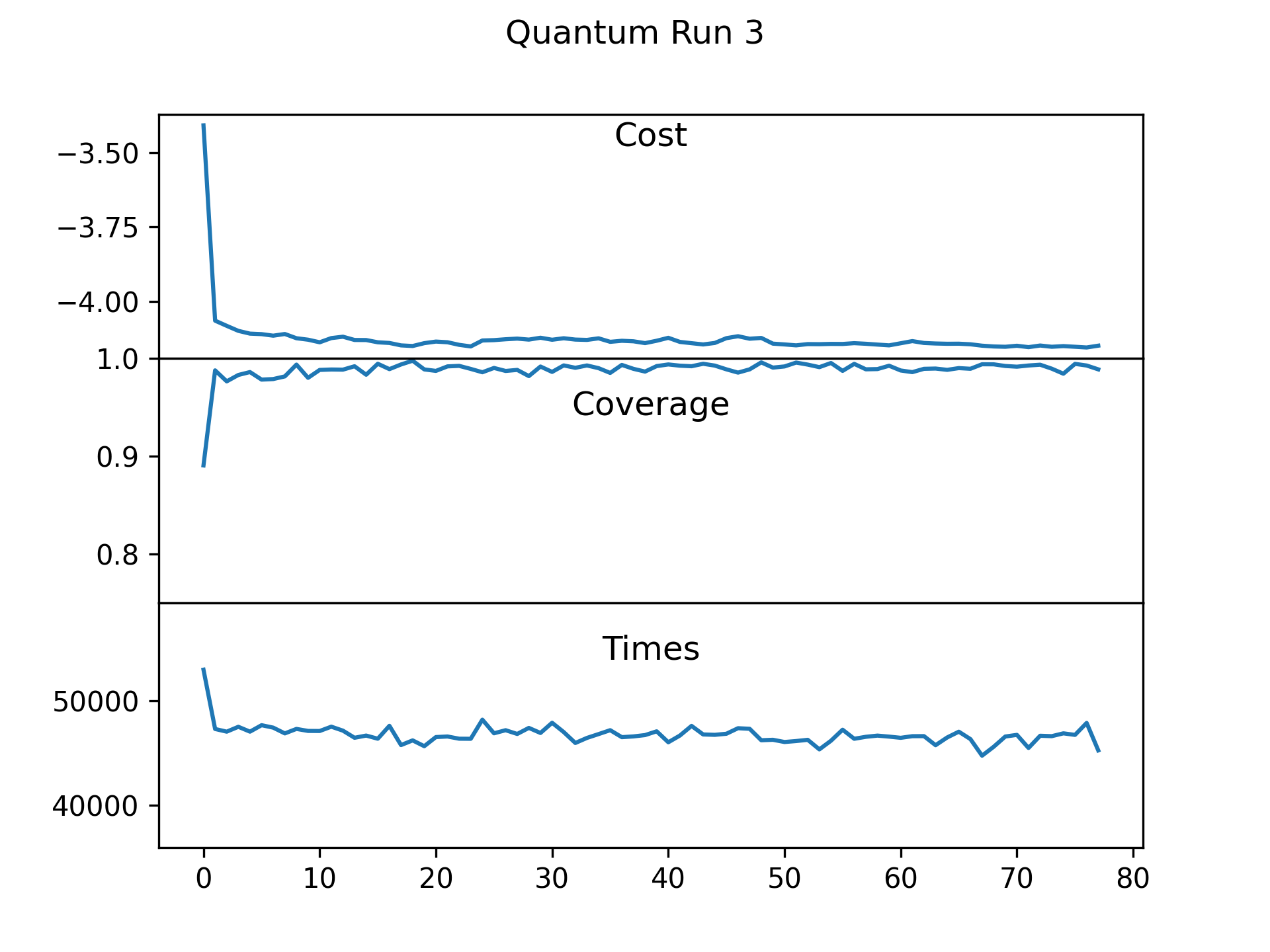}
\caption{Quantum training run.  2 truck, rank 2 and rank 3 demand, with
no cyclic box-return constraint. Uses 8 nodes.}
\label{fig:quantum3}
\end{figure}

Next we compare a simple scenario (similar to Quantum Run 1) when including the cyclic demand structure or not.  First, without the cyclic demand, figure \ref{fig:quantum3} shows similar results as before for the quantum training results versus training epoch.

\begin{figure}
\centering
\includegraphics[width=.5\textwidth]{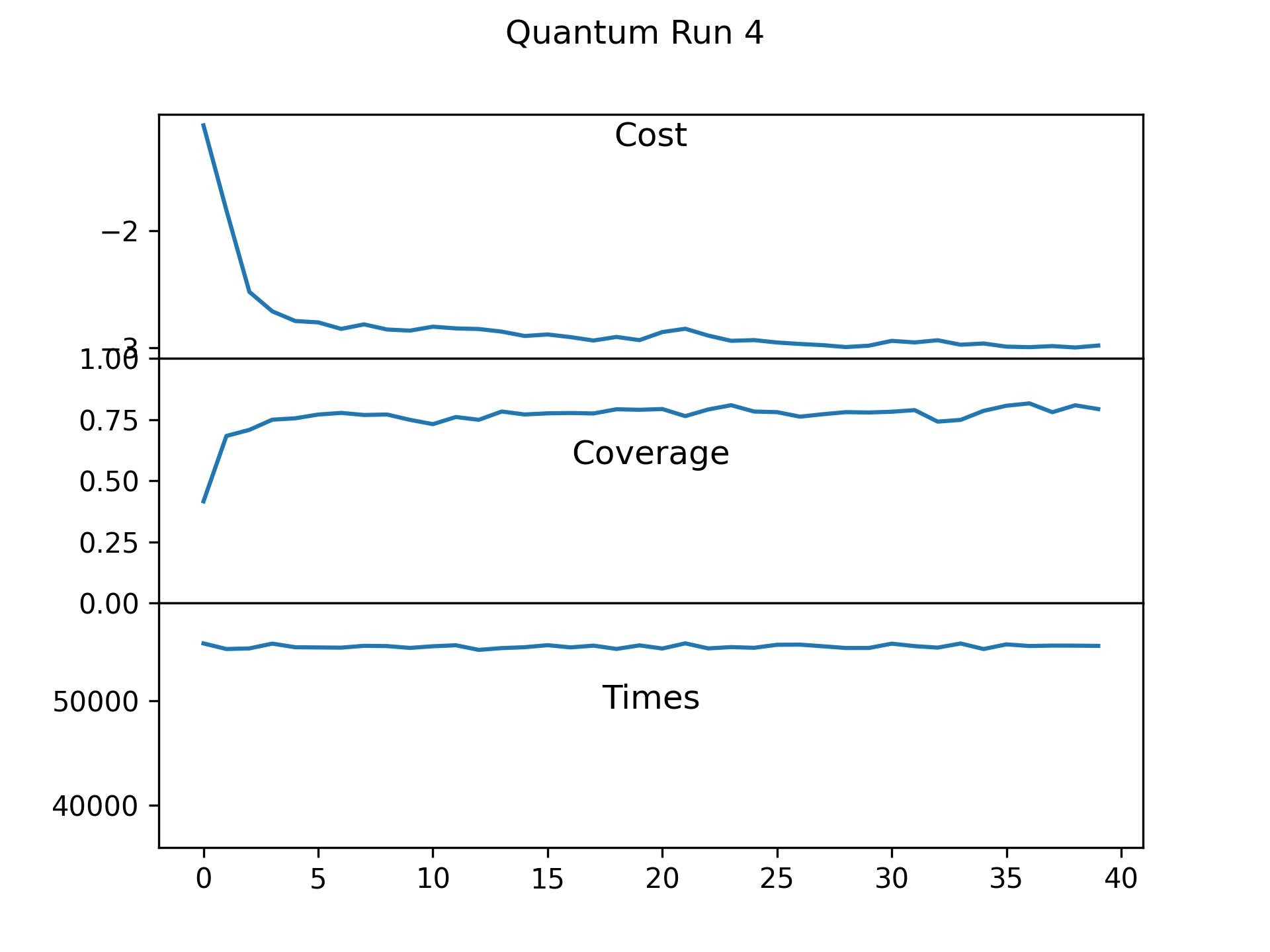}
\caption{Quantum training run.  2 truck, rank 2 cyclic demand. Uses 8 nodes.}
\label{fig:quantum4}
\end{figure}

And then we see that adding the more realistic cyclic demand structure degrades the training performance.  Figure \ref{fig:quantum4} shows the quantum training results versus training epoch for the same problem as before, but now with the more realistic cyclic rank 2 demand structure.

These results show that the quantum training is challenged to include the more realistic problem features, such as rank 3 demand structure and the cyclic box return constraint.  This will limit what we can implement in quantum hardware at this time regarding quantum training.

\subsection{Results: AVRP Solutions with Ion Trap Quantum Hardware}
\label{sec:ionq-solutions}

\begin{figure}
\centering
\includegraphics[width=.5\textwidth]{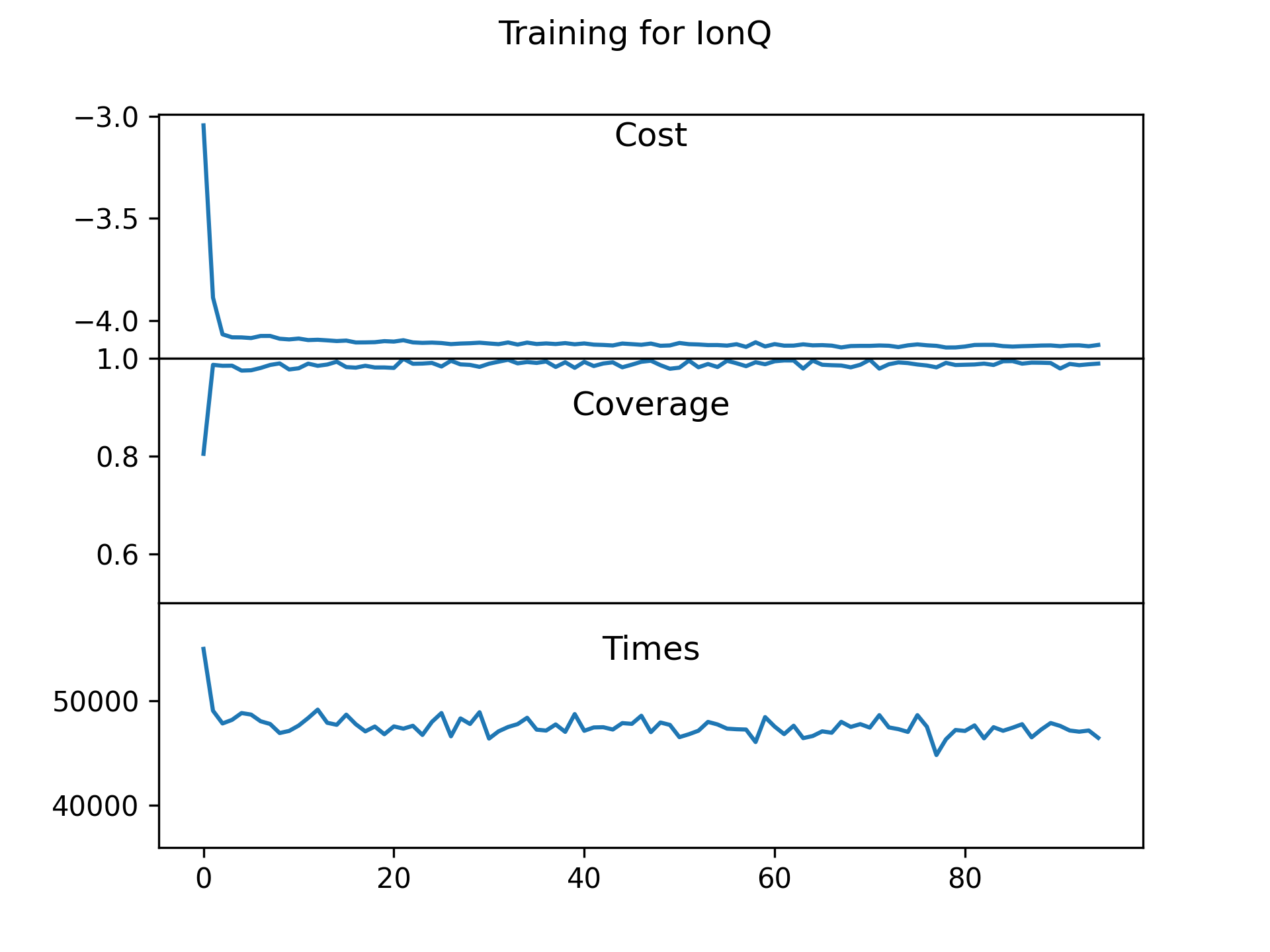}
\caption{Classical training run.  2 truck, rank 2 only demand, with
no cyclic box-return constraint. Uses 8 nodes.}
\label{fig:ionq-training}
\end{figure}

And finally we test how well the quantum circuit computes solution inferences.  To do this, we first make the choice to use a classically trained model that can capture more of the complexity of the problem.   The approach for executing on IonQ is to use
a previously trained reinforcement learning
agent (see figure \ref{fig:ionq-training}. The agent was trained using QC Ware's
quantum circuit emulator, but for execution
we use IonQ. For every key in two encoder
layers, there are 8 attention heads. We use
an encoding dimension of 64, meaning that
each of the 8 attention head keys has a
dimension of 8 itself. We pass each
key through an 8 qubit quantum ONN.
Moreover, there are 8 nodes. This
means that we have 
\begin{equation*}
    \underset{\text{nodes}}{8}\times\underset{\text{heads}}{8}\times\underset{\text{encoding layers}}{2}\times\underset{\text{tomography steps}}{3}=384\text{ circuits}
\end{equation*}
Each of these circuits is run with 500 measurements.

\subsubsection*{Results of IonQ Run}
Our two-truck IonQ execution on 8 nodes with simplified
demand resulted in the following route selection:

\begin{center}
\begin{tabular}{ c }
\tiny
\begin{lstlisting}

      Truck  Departure Time              Departure Node
0   Truck 0               0                  KIRA PLANT
1   Truck 1               0                       MEIKO
2   Truck 0            3960            NO.1 AND 2 PLANT
3   Truck 1            6300        NISHIO CROSS-DOCKING
4   Truck 0            7200        NISHIO CROSS-DOCKING
5   Truck 1           11280          OKAZAKI EAST PLANT
6   Truck 0           15120                TAHARA PLANT
7   Truck 1           16380            NO.1 AND 2 PLANT
8   Truck 1           19620        NISHIO CROSS-DOCKING
9   Truck 0           23040        NISHIO CROSS-DOCKING
10  Truck 1           25860                       MEIKO
11  Truck 0           28020          OKAZAKI EAST PLANT
12  Truck 1           31500          OKAZAKI EAST PLANT
13  Truck 0           33060        NISHIO CROSS-DOCKING
14  Truck 1           36540        NISHIO CROSS-DOCKING
15  Truck 0           37560  OKAZAKI AND ELECTRIC PLANT
16  Truck 1           41040  OKAZAKI AND ELECTRIC PLANT
17  Truck 0           42600              GAMAGORI PLANT
18  Truck 1           48420                TAHARA PLANT
19  Truck 0           49140                TAHARA PLANT
20  Truck 1           55020              GAMAGORI PLANT
21  Truck 0           55740              GAMAGORI PLANT

\end{lstlisting}
\normalsize
\end{tabular}
\end{center}

\begin{figure}
\centering
\includegraphics[width=.5\textwidth]{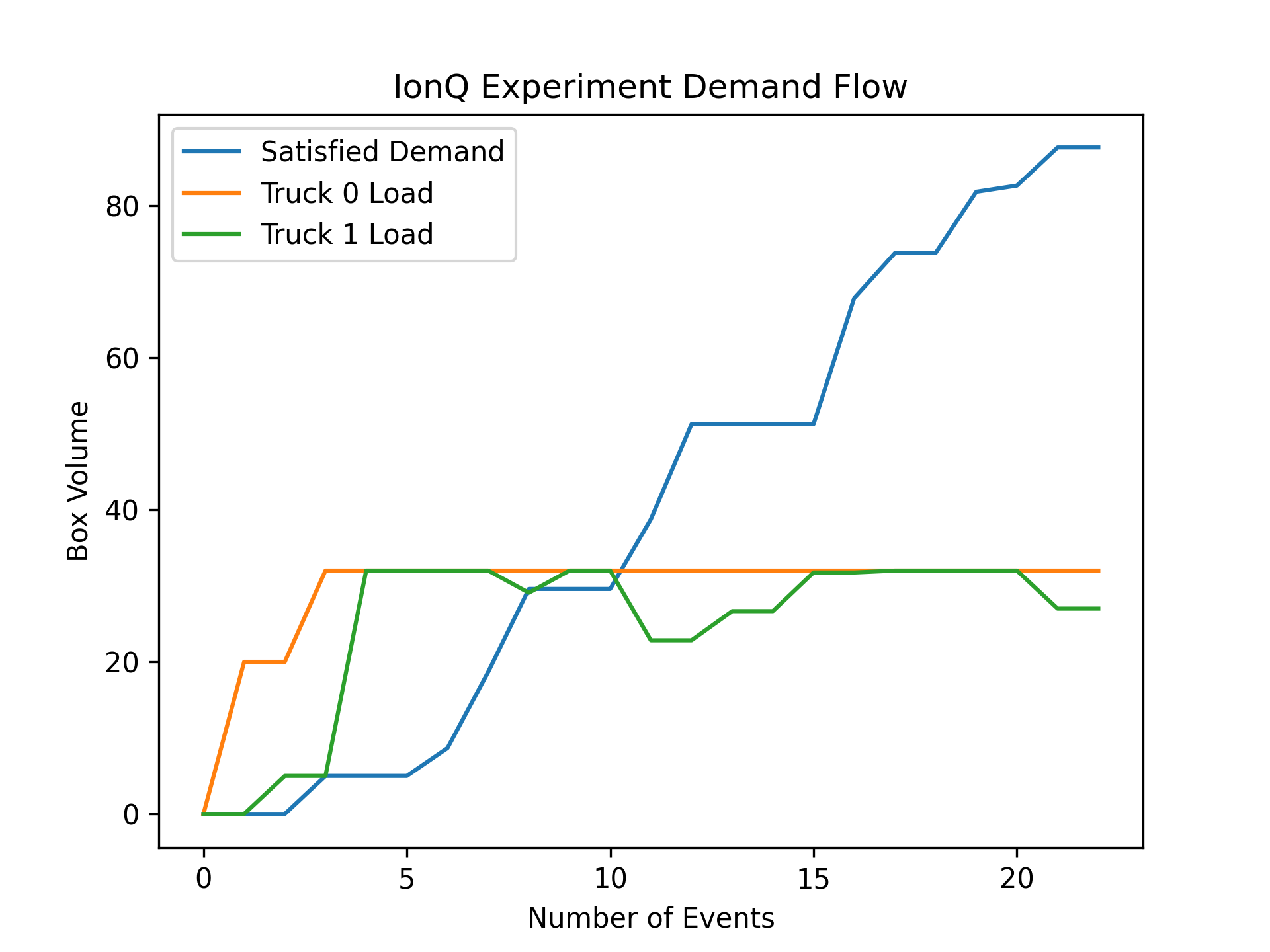}
\caption{Demand share for IonQ truck assignments. The 
total initial rank 2 direct demand was 179, so we see
that the truck successfully satisfy 49\% of demand. This is not in conflict training data \ref{fig:ionq-training}
because for that training data we clipped to a very small
amount of demand in each training episode.}
\label{fig:ionq-sat}
\end{figure}

The quality of the route found is illustrated
by figure \ref{fig:ionq-sat}. This figure
shows the volume carried by trucks at various times
as well as the total satisfied demand. While
the initial demand of 179 is only 49\% satsified,
this percentage is actually much better than it sounds.
The training data \ref{fig:ionq-training} was done
with data that was clipped to very small demand
to make training manageable. However, the actual
run used a much higher demand scale so there is
no reason to regard 49\% as low coverage. 

In principle, we could cover the full Aisin supply chain
with groups of two trucks, each computed with
IonQ. However, the financial cost of this would be
enormous. The run we discuss in this section alone 
incurred over USD 2000 to run, and this does not
account for costs of additional experiments for 
obtaining improved data.

\section{Conclusions}

While there is much additional work to do in exploring algorithmic and workflow improvements, these first results demonstrate that quantum circuits can successfully train a reinforcement model comprised of ONNs to solve the logistics problem.  While the quantum versions of ONNs, QONNs, have theoretically provable speedups, it remains to be seen how much of that theoretical advantage can be realized in practice.  

To that end, this work demonstrated the successful implementation of QONNs applied to a practical problem of real world importance and developed several methods of tuning and measuring performance, both of the quantum circuit primitives used in the QONN and in the overall metaparameters used in constructing the training and inference workflows.

We were only able to study a few approaches of implementing the QONN, quantum circuits, attention heads, layering depth, demand structure, and constraints during the course of this project.  Further work would likely find more efficient circuits and QONN implementations along with more effective training approaches.  

We emphasize that the errors in quantum computing we are seeing here are
not surprising and reflect the early state of quantum hardware. So it will be important in the future to continue monitoring the improvements in quantum hardware and the performance on algorithmic primitives to discern when they might improve computational workflows.

 \bibliographystyle{unsrtnat}
 \bibliography{bibliography.bib}

\end{document}